\documentclass[aip,cha,preprint,numerical]{revtex4-2}
\usepackage{amsmath}
\usepackage{amssymb}
\usepackage{graphicx}
\usepackage{natbib}		
\usepackage{color}

\def\vp{\varphi}
\def\w{\omega}
\newcommand{\ii}{\mathrm{i}}

\newcommand{\av}[1]{\left\langle #1 \right\rangle}

\begin{document}

\title{Dynamics of large oscillator populations with random interactions}
\author{Arkady Pikovsky}
\affiliation{Institute of Physics and Astronomy, University of Potsdam,
Karl-Liebknecht-Str. 24/25, 14476 Potsdam-Golm, Germany}
\author{Lev A. Smirnov}	 
\affiliation{Department of Control Theory, Lobachevsky State University of Nizhny Novgorod,
23 Gagarin Avenue, Nizhny Novgorod 603022, Russia}

\date{\today}

\begin{abstract}
We explore large populations of phase oscillators interacting via random coupling functions. Two types of coupling terms, the Kuramoto-Daido coupling and the Winfree coupling, are considered. Under the assumption of statistical independence of the phases and the couplings, we derive reduced averaged equations with effective non-random coupling terms. As a particular example, we study interactions that have the same shape but possess random coupling strengths and random phase shifts. While randomness in coupling strengths just renormalizes the interaction, a distribution of the phase shifts in coupling reshapes the coupling function.
\end{abstract}

\maketitle

\begin{quotation}
Populations of globally coupled oscillators appear in different fields of physics, engineering, and life sciences. In many situations, there is disorder in the coupling, and the coupling terms are not identical but vary, for example, due to different phase shifts and coupling strengths. We show that for large ensembles, one can effectively average over the disorder and describe the synchronization transition via simpler averaged equations. 
\end{quotation}

\section{Introduction}
\label{sec:intro}

Collective synchronization in oscillator populations attracted much interest in the last decades~\cite{Winfree-67,Kuramoto-75,Strogatz-Stewart-93,Pikovsky-Rosenblum-Kurths-01}. While the phenomenon is well-understood in a regular situation, the influence of disorder remains a subject of intensive current studies. 
The disordered case is relevant for many applications, especially in neuroscience, where
in the description of the correlated activity of neurons~\cite{Glass-Mackey-88,Buzsaki-06,daffertshofer2020phase}, one can hardly assume the neurons themselves to be identical and the coupling between them to be uniform.

Randomness in the form of quenched (time-independent) disorder can appear in different forms. Already in the original paper by Kuramoto~\cite{Kuramoto-75}, a distribution of the natural frequencies of oscillators was taken into account. In further studies, effects of pairwise disorder in couplings have been formulated by a random matrix of coupling strengths~\cite{kalloniatis2010incoherence,chiba2018bifurcations,juhasz2019critical}. In this paper, we consider another type of disorder in interactions, which is maximal in some sense: we assume that all the pairwise coupling terms are different, taken from some random distribution of random functions. 
In particular, the coupling functions can have different shapes: some possess just one
harmonic of the phases or of the phase differences, and some can be more complex with many harmonics.
Such a setup accounts for a maximal heterogeneity of coupled units.   

In numerical examples in this paper, we restrict to a subclass where the coupling functions have the same shape but differ due to random
coupling strengths and to random phase shifts in the coupling~\cite{Park-Rhee-Choi-98}. Phase shifts in the coupling lead to frustration, which can be understood as follows. Synchronization is always possible for two coupled oscillators with different phase shifts in the coupling terms.  In this synchronous regime, the frequencies of the oscillators coincide, but the phases are not equal; there is some difference in the phases dependent on the phase shifts in the coupling. If many oscillators interact with different phase shifts, optimal phase relations in pairs may become conflicting, and the main problem is how this frustration affects global synchrony. In a recent short communication~\cite{smirnov2024dynamics}, we argued that in the thermodynamic limit of many coupled oscillators, one can perform averaging over random phase shifts, which results in a new effective coupling function. This paper extends this approach to different setups and illustrates it with numerical simulations.

The paper is organized as follows. In Section~\ref{sec:setups}, we formulate different setups for the phase dynamics of coupled oscillators and rotators. In Section~\ref{sec:mrand}
we present our main theoretical result about reducing the disordered situation to an effective regular averaged coupling function. A particular situation where randomness is in the coupling strengths and in the phase shifts is considered in Section~\ref{sec:rcsps}.
The validity of this reduction is illustrated by numerical examples in Section~\ref{sec:ne}. In Section~\ref{sec:gps}, we show how to treat systems with partial disorder, where random phase shifts are attributed to a driving or to a driven unit. We conclude with a discussion in Section~\ref{sec:conc}.

\section{Different regular setups for coupled phase oscillators}
\label{sec:setups}
This section describes several popular models of interacting phase oscillators without randomness in coupling. For our purposes, it is important to separate the individual dynamics and the coupling terms, which we denote ``c.t.'' .

\subsection{Winfree and Kuramoto-Daido coupling terms}

The ``standard'' model describes the individual phase dynamics of an oscillator as rotations with a natural frequency $\w_k$, possibly with individual noises $\sigma \xi_k(t)$:
\begin{equation}
\dot\vp_k=\w_k+\sigma \xi_k(t)+\text{c.t.}\;.
\label{eq:sm}
\end{equation} 
The model \eqref{eq:sm} is most popular because it can be directly derived for generic coupled oscillators. However, in the literature, several similar particular setups have been discussed. The model of coupled active rotators~\cite{Shinomoto-Kuramoto-86,Sakaguchi-Shinomoto-Kuramoto-88,Klinshov2021noise} includes a non-uniform rotation of the variable $\vp_{k}$ (which is thus not interpreted as the oscillator phase):
\begin{equation}
\dot\vp_k=\w_k-\nu\sin\vp_k+\sigma \xi_k(t)+\text{c.t.}\,.
\label{eq:ar}
\end{equation}
Another case is where $\vp_{k}$ is an angle variable (or the difference of the phases of macroscopic wave functions for Josephson junctions) with a second-order in time dynamics~\cite{Tanaka-Lichtenberg-Oishi-97,Hong-Choi-Yi-Soh-99,Munyaev_etal_2020,Munyayev_etal_2023}
(recently, this model became popular in the context of modeling power grids~\cite{Filatrella_etall-08,dorfler2012synchronization}):
\begin{equation}
\mu\ddot{\vp}_k+\dot\vp_k=\w_k+\sigma \xi_k(t)+\text{c.t.}\,.
\label{eq:rm}
\end{equation}

Next, we specify the coupling terms ``c.t.'' according to the Winfree and the Kuramoto-Daido approaches. 
In the Winfree model, the action on the oscillator $k$ from the oscillator $j$ is proportional to the product $S(\vp_k)Q(\vp_j)$, where $S(\vp_k)$ is the phase sensitivity function of the unit $k$, and $Q(\vp_j)$ describes the force with which the element $j$ is acting. The latter is usually renormalized according to the number $N$ of interacting oscillators to have a well-defined thermodynamic limit. The full Winfree-type coupling term reads~\cite{Winfree-67,Kuramoto-84,Pikovsky-Rosenblum-Kurths-01}
\begin{equation}
\frac{1}{N}S(\vp_k)\sum_j Q(\vp_j)\,.
\label{eq:wm}
\end{equation}
The Winfree model can be directly derived from the original equations governing the oscillator dynamics, in the first order in the small parameter describing the coupling strength~\cite{Kuramoto-84,Pikovsky-Rosenblum-Kurths-01}.

Further usage of this small parameter allows for a reduction of the Winfree model \eqref{eq:wm} to the Kuramoto-Daido model. Because the coupling is weak, the phase dynamics is represented by a fast rotation with frequency $\w_k$ and a slow variation due to the coupling. Because only the slow dynamics is responsible for large deviations of the phases, one can perform averaging over fast oscillations, keeping only the components with slow phase dependencies in the coupling term. If the frequencies of the oscillators are nearly equal, then the slow combinations of the phases are $\sim(\vp_j-\vp_k)$. Thus, keeping only slow terms yields the coupling term in the Kuramoto-Daido form
\begin{equation}
\frac{1}{N}\sum_j F(\vp_j-\vp_k)\,.
\label{eq:kd}
\end{equation}

\subsection{Formulation in terms of Kuramoto-Daido order parameters}
Here, we show how the Kuramoto-Daido and the Winfree coupling functions can be reformulated in terms of the Kuramoto-Daido order parameters. The latter are defined as
\begin{equation}
Z_m=\av{e^{\ii m\vp_{j}}}=\frac{1}{N}\sum_j e^{\ii m\vp_j}\,.
\label{eq:kop}
\end{equation}
We start with the Kuramoto-Daido-type coupling~\eqref{eq:kd} and represent the $2\pi$-periodic coupling function $F(x)$ as a Fourier series
\begin{equation}
F(x)=\sum_m f_m e^{\ii mx}\,,\qquad f_m=\!\frac{1}{2\pi}\int_0^{2\pi}\!\!dx\,F(x)e^{-\ii mx}\,.
\label{eq:ff}
\end{equation}
Substituting this in the coupling term in \eqref{eq:kd} we obtain, using expression \eqref{eq:kop},
\begin{equation}
\frac{1}{N}\sum_jF(\vp_j-\vp_k)=\sum_m f_m e^{-\ii m\vp_k}\frac{1}{N}\sum_j e^{\ii m\vp_j}=\sum_m f_m e^{-\ii m\vp_k} Z_m\;.
\label{eq:kdfm}
\end{equation}
Similarly, for the Winfree case~\eqref{eq:wm} we can represent the coupling functions $Q(x)$ and $S(x)$ as Fourier series
\begin{equation}
\begin{gathered}
S(x)\!=\!\sum_m s_m e^{\ii mx}\,,\qquad
s_m\!=\!\frac{1}{2\pi}\int_0^{2\pi}\!\!dx\,S(x)e^{-\ii mx}\,,\\
Q(x)=\sum_l q_l \,e^{\ii lx}\,,\qquad
q_l=\frac{1}{2\pi}\int_0^{2\pi}\!\!dx\,Q(x)\,e^{-\ii lx}\,.\label{eq:w2}
\end{gathered}
\end{equation}
Substitution in Eq.~\eqref{eq:wm} allows for a representation of the coupling term as
 \begin{equation}
\frac{1}{N}\sum_j \sum_m s_m e^{\ii m\vp_k} \sum_l q_l e^{\ii l\vp_j}\!=\!
\sum_m\sum_l s_m e^{\ii m\vp_k} \,q_l \left[ \frac{1}{N}\sum_j e^{\ii l\vp_j} 
\right]\!=\!\sum_m s_m e^{\ii m\vp_k}\sum_l q_l Z_l \;.
\label{eq:winop}
\end{equation}
Below, we will use expressions \eqref{eq:kdfm} and \eqref{eq:winop} as ``templates'' for identifying the effective coupling functions in the case of random interactions.

\section{General random coupling functions}
\label{sec:mrand}

We start with the maximally disordered setup, where all pairwise coupling functions are random.
We consider first the Kuramoto-Daido-type coupling \eqref{eq:kd} and rewrite it assuming that all the coupling terms are generally different 
\begin{equation}
\frac{1}{N}\sum_j F_{jk}(\vp_j-\vp_k)\,.
\label{eq:kdmr}
\end{equation}
We use the Fourier representation like \eqref{eq:ff}:
\begin{equation}
F_{jk}(x)=\sum_m f_{m,jk} e^{\ii mx}\;,\qquad f_{m,jk}=\frac{1}{2\pi}\int_0^{2\pi}dx F_{jk}(x)e^{-\ii mx}\,.
\label{eq:ffmr}
\end{equation}
Here, we treat the complex Fourier coefficients $f_{m,jk}$ as random numbers with some distribution. 
The randomness in coupling is a quenched (i.e., time-independent) disorder.
 
We now perform the averaging of the coupling \eqref{eq:kdmr}. First, we substitute \eqref{eq:ffmr} in \eqref{eq:kdmr}
\begin{equation}
\frac{1}{N}\sum_j \sum_m f_{m,jk}e^{\ii(m\vp_j-m\vp_k)}=\sum_m  e^{-\ii m\vp_k} \left[
\frac{1}{N}\sum_j f_{m,jk} e^{\ii m\vp_j}\right].
\label{eq:kdmr1}
\end{equation}  
We identify the term in the squared brackets as the population 
average $\av{f_{m,jk} e^{\ii m\vp_{j}}}$. Next, we assume statistical independence of the phases and the Fourier coefficients $f_{m,jk}$. We expect this independence to be valid for a large population, where many different couplings influence each phase. This assumption allows us for representing this average as $\av{f_{m,jk}}\av{e^{\ii m\vp_{j}}}=\av{f_{m,jk}}Z_m$. As a result, we obtain the coupling term
\begin{equation}
\sum_m \av{f_{m,jk}} e^{-\ii m\vp_k }Z_m\,.
\label{eq:kdmr2}
\end{equation}
Comparing this expression with \eqref{eq:kdfm}, we conclude that the coupling is described with an effective deterministic coupling function, Fourier modes of which are just $\av{f_{m,jk}}$. 

In the case of the Winfree-type coupling \eqref{eq:wm}, the case of general randomness is one with all different phase sensitivity functions $S_{jk}(x)$ and forcing functions $Q_{jk}(x)$:
\begin{equation}
\frac{1}{N}\sum_j S_{jk}(\vp_k) Q_{jk}(\vp_j)\;.
\label{eq:wmmr}
\end{equation}
Again, we represent these functions via random complex Fourier coefficients $s_{m,jk}$ and $q_{l,jk}$ 
\begin{equation}
\begin{gathered}
S_{jk}(x)\!=\!\sum_m s_{m,jk} e^{\ii mx}\,,\qquad
s_{m,jk}\!=\!\frac{1}{2\pi}\int_0^{2\pi}\!\!dx\,S_{jk}(x)e^{-\ii mx}\,,\\
Q_{jk}(x)=\sum_l q_{l,jk}\, e^{\ii lx}\,,\qquad
q_{l,jk}=\frac{1}{2\pi}\int_0^{2\pi}\!\!dx\,Q_{jk}(x)\,e^{-\ii lx}\,.\label{eq:w2mr}
\end{gathered}
\end{equation}
Substituting \eqref{eq:w2mr} in \eqref{eq:wmmr} and assuming statistical independence of the phases and the Fourier coefficients, we arrive at an effective deterministic  coupling
\begin{equation}
\sum_m \av{s_{m,jk}} e^{\ii m\vp_k} \sum_l \av{q_{l,jk}} Z_l\;. 
\label{eq:wmmr1}
\end{equation}
Comparison of this expression with \eqref{eq:winop} shows that the Fourier modes of the effective phase sensitivity function and of the forcing are just the averaged random Fourier modes.

The last remark is that because the Fourier transform is a linear operation, averaging the Fourier coefficients is the same as averaging the functions. Thus, the effective averaged coupling function in the Kuramoto-Daido case \eqref{eq:kdmr} is
\begin{equation}
\frac{1}{N}\sum_j F_{jk}(\vp_j-\vp_k)\;\Rightarrow\;
\frac{1}{N}\sum_j \mathcal{F}(\vp_j-\vp_k)=\frac{1}{N}\sum_j \av{F_{jk}(\vp_j-\vp_k)}\;.
\label{eq:kdav}
\end{equation}   
For the random Winfree coupling \eqref{eq:wmmr}, we have
\begin{equation}
\begin{gathered}
\frac{1}{N}\sum_j S_{jk}(\vp_k) Q_{jk}(\vp_j)\;\Rightarrow\;
\mathcal{S}(\vp_k)\frac{1}{N}\sum_j \mathcal{Q}(\vp_j)\;,\\
\mathcal{S}(\vp_k)=\av{S_{jk}(\vp_k)},\quad \mathcal{Q}(\vp_j)=\av{Q_{jk}(\vp_j)}\;.
\end{gathered}
\label{eq:wmav}
\end{equation}

\section{Randomness in coupling strengths and in the phase shifts}
\label{sec:rcsps}

The case of general randomness of interactions, presented in Section~\ref{sec:mrand}, includes a situation where different interactions have different shapes. For example, some oscillators can be coupled via the first harmonic coupling function, while others are coupled with the second harmonic coupling function. A particular situation is one where all the shapes are the same, but the interactions differ in their phase shifts and the coupling strengths. For the Kuramoto-Daido couplings \eqref{eq:kdmr}, this means that 
\begin{equation}
F_{jk}(\vp_j-\vp_k)= A_{jk} F(\vp_j-\vp_k-\alpha_{jk})\,.
\label{eq:kd-pc}
\end{equation}
Here, random numbers $A_{jk}$ describe different random coupling strengths, and $\alpha_{jk}$ are random phase shifts. For simplicity of presentation, we assume statistical independence of $A_{jk}$ and $\alpha_{jk}$. Then, the random coupling \eqref{eq:kd-pc} is described by two distributions. We denote the distribution of the coupling strengths $U(A)$ and the distribution of the phase shifts $G(\alpha)$. This latter distribution is defined on a circle and can be represented by the Fourier series:
\begin{equation}
G(\alpha)=\frac{1}{2\pi}\sum_m g_m e^{\ii m\alpha}\,,\quad g_m=\!\int_0^{2\pi}\!\!d\alpha\,G(\alpha)e^{-\ii m\alpha}\,.
\label{eq:adist}
\end{equation}
Such pairwise phase shifts naturally appear if there are time delays for the pairwise couplings.
As has been demonstrated in Ref.~\onlinecite{Izhikevich-98}, a phase shift $\alpha$ is determined in the leading order in coupling strength as follows: $\alpha=\w \tau$,
where $\tau$ is the signal propagation time and $\w$ is the characteristic frequency (e.g., a median frequency if the natural frequencies are different).
Remarkably, phase shifts like in~\eqref{eq:kd-pc} naturally appear in the description of power grids (cf. Eq. (3) in Ref.~\onlinecite{dorfler2012synchronization}).

Applying general expression \eqref{eq:kdav} to this particular case, we obtain the effective
coupling function as
\begin{equation}
\mathcal{F}(x)=\av{A}\int_0^{2\pi} F(x-\alpha) G(\alpha)\;d\alpha\;.
\label{eq:kmampps}
\end{equation}
where $\av{A}=\int dA\, A\,U(A)$.
One can see that the randomness of coupling strengths renormalizes the total coupling strength, but does not influence the shape of the coupling function. In contradistinction, the randomness of the phase shifts changes the form of the coupling function because of the convolution operator in \eqref{eq:kmampps}. The best way to see this is to return to the Fourier representation, where the effect of the randomness in the phase shifts reduces to a factorization of the Fourier modes
\begin{equation}
f_m\;\Rightarrow\; f_m g_m\;.
\label{eq:kdfreq}
\end{equation}
One can see that some modes in the coupling function can even disappear if the corresponding factors $g_m$ vanish. We will explore such cases in Section~\ref{sec:ne} below.

In the context of the Winfree model \eqref{eq:wm}, one can also restrict randomness to coupling strengths and phase shifts so that the shapes of the phase sensitivity function and of the forcing remain the same.  We have, in general, two phase shifts $\alpha_{jk}$ and $\beta_{jk}$, entering the phase sensitivity function and the driving term, respectively, and coupling strengths $A_{jk}$ (for simplicity of presentation we consider all the three random quantities as statistically independent):
\begin{equation}
\frac{1}{N}\sum_j A_{jk} S(\vp_k-\alpha_{jk}) Q(\vp_j-\beta_{jk})\,.
\label{eq:wmp1}
\end{equation}
Of course, one of these phase shifts may be absent in a particular situation.
The distribution $B(\beta)$ of phase shifts $\beta$ is represented similarly to \eqref{eq:adist}:
\begin{equation}
B(\beta)=\frac{1}{2\pi}\sum_m b_m e^{\ii m\beta},\quad b_m=\int_0^{2\pi}d\beta\,B(\beta)e^{-\ii m\beta}\,.
\label{eq:bdist}
\end{equation}
Now, we apply general expressions \eqref{eq:wmav} to the case of randomness in phase shifts and coupling strengths and obtain:
\begin{equation}
\begin{gathered}
\mathcal{S}(x)=\av{A}\av{S(x-\alpha_{jk})}=\av{A}\int_0^{2\pi}S(x-\alpha) G(\alpha)\;d\alpha \;,\\ \mathcal{Q}(y)=\av{Q(y-\beta_{jk})}=\int_0^{2\pi}Q(y-\beta) B(\beta)\;d\beta \;.
\end{gathered}
\label{eq:wmav2}
\end{equation}
Here, we, somewhat arbitrarily,  attributed the average coupling strength to the phase sensitivity function. Again, like in the case of the Kuramoto-Daido-type coupling, the coupling strengths do not influence the shape of the functions, while the distributions of the phase shifts do reshape them. In terms of Fourier modes, one has a factorization by the modes of the distributions: 
\begin{equation}
s_m\;\Rightarrow\; s_m g_m\;,\qquad q_l\;\Rightarrow q_l b_l\;.
\label{eq:winfm}
\end{equation}

\section{Numerical examples}
\label{sec:ne}

The theory above predicts that in the presence of random phase shifts in the coupling, the effective coupling function is the convolution of the original one with the phase shift distribution density. In terms of the Fourier modes, one has a product of the modes of the original coupling function with the Fourier modes of the distribution density of the phase shifts. The most prominent effect appears if the original coupling function is rather complex (has several harmonics), but the distribution of the phase shifts is simple, possessing just one harmonic. For example, we take the density of the phase shifts $G(\alpha)$ in the form ($M$ is an integer)
\begin{equation}
G(\alpha)=\frac{1}{2\pi}(1+\cos M \alpha)\;.
\label{eq:psd}
\end{equation}
In the Fourier representation, this corresponds to one nontrivial Fourier mode $g_M=0.5$. Correspondingly, only the $M$-th harmonic is present in the effective coupling function with such a distribution of the phase shifts.

\begin{figure}[!t]
\centering
\includegraphics[width=0.5\columnwidth]{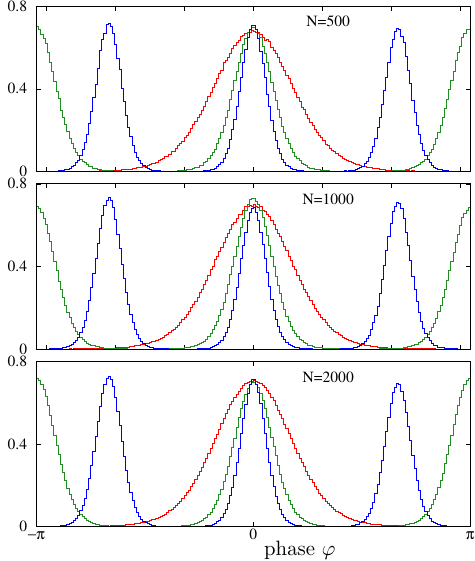}
\caption{Estimations of the phase probability densities via histograms (200 bins are used)  in model \eqref{eq:ex1},~\eqref{eq:psd} with $M=1$ (red lines), $M=2$ (green lines), and $M=3$ (blue lines). These distributions are almost the same for different system sizes $N$.}
\label{fig:ex1}
\end{figure}

In Fig.~\ref{fig:ex1}, we show the distributions of the phases in globally coupled populations of noisy identical phase oscillators ($\omega_{k}=0$) with the Kuramoto-Daido coupling
\begin{equation}
\dot\vp_k=\sigma\xi_k(t)+\frac{1}{N}\sum_j F(\vp_j-\vp_k-\alpha_{jk}),\quad 
F(x)=\sin(x)+2\sin(2x)+3\sin(3x)\;.
\label{eq:ex1}
\end{equation}
We show the results for $\sigma=0.12$ and three distributions of type \eqref{eq:psd} with $M=1,2,3$.
One can see that in these cases, the densities possess the corresponding $(2\pi/M)$-periodicities as functions of $\vp$,
confirming that the phase shifts result in the effective one-mode coupling.

\begin{figure}[!t]
\centering
\includegraphics[width=0.75\columnwidth]{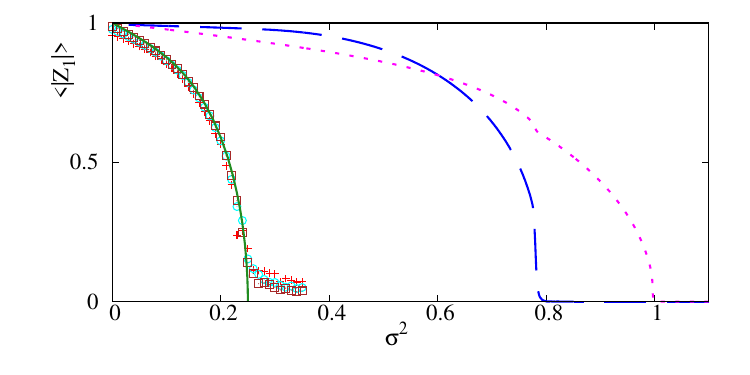} 
\caption{Dashed and dotted lines (order parameters $\av{\left|Z_{1}\right|}=\av{\left|\av{e^{\ii\vp_{j}}}\right|}$
and $\av{\left|Z_{2}\right|}=\av{\left|\av{e^{2\ii\vp_{j}}}\right|}$, respectively):
the behavior of an ensemble of noisy identical oscillators with Kuramoto-Daido coupling in the form~\eqref{eq:2h}
with $K=1$ in dependence on the noise intensity $\sigma^{2}$ in the absence of phase shifts in coupling.
Markers: order parameter $\av{\left|Z_{1}\right|}$ in simulation of a disordered population with the phase shifts $\alpha_{jk}$
sampled according to~\eqref{eq:psd} with $M=1$ ($N=500$: red pluses; $N=1000$: cyan circles, $N=2000$: brown squares).
Here and below, we also apply time averaging when calculating complex mean fields to smooth out small-magnitude, non-regular fluctuations caused by noise and finite-size effects.
Solid green line: theoretical prediction of the first order parameter $\left|Z_{1}\right|$ in the reduced model according to Ref.~\onlinecite{Munyaev_etal_2020}.}
\label{fig:ex2}
\end{figure}

\begin{figure}[!t]
\centering
\includegraphics[width=0.645\columnwidth]{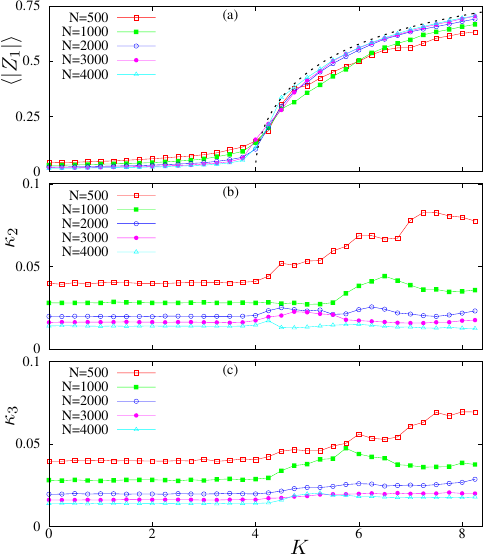} 
\caption{Panel (a): order parameter $\av{\left|Z_{1}\right|}$ for ensembles with random phase shifts $\alpha_{jk}$
taken from the one-mode distribution~\eqref{eq:psd} with $M=1$, random natural frequencies $\omega_{k}$ drawn
from the Cauchy distribution~\eqref{eq:cauchy} with $\Delta=1$, and coupling function~\eqref{eq:2h}.
Dashed line: theoretical prediction of the order parameter according to Ref.~\onlinecite{Kuramoto-75}: $|Z_{1}|=\sqrt{(K-4)\bigl/K\bigr.}$.
Panels (b,\,c): averaged over time absolute values of the second and third circular cumulants $\kappa_{2}\!=\!\av{\left|Z_{2}-Z_{1}^{2}\right|}$
and $\kappa_{3}\!=\!\av{\left|Z_{3}-3Z_{2}Z_{1}+2Z_{1}^{3}\right|}$, respectively;
their small values confirm that the distribution of the phases is a wrapped Cauchy distribution.}
\label{fig:ex3}
\end{figure}

A more quantitative test can be performed if an exact analytical solution exists for the reduced system.
Such solutions are known for the Kuramoto-Daido system with only the first harmonic in the coupling function,
in the cases of noisy identical oscillators~\cite{Munyaev_etal_2020} and for deterministic oscillators with a Cauchy distribution of natural frequencies~\cite{Kuramoto-75}:
\begin{equation}
W(\w)=W_{C}(\w-\Omega), \qquad W_{C}(\varpi)=\Delta\Bigl/\pi\left(\varpi^{2}+\Delta^{2}\right)\Bigr.\,,
\label{eq:cauchy}
\end{equation}
where $\Omega$ is the mean value, and $\Delta$ is the parameter governing the width of the distribution.
In the next two examples, we fix $M=1$ in~\eqref{eq:psd} and take a two-harmonic original coupling function 
\begin{equation}
F(x)=K\bigl(\sin(x)+4\sin(2x)\bigr)\,.
\label{eq:2h}
\end{equation}
The effective coupling function is thus $\mathcal{F}(x)\!=\!0.5\,K\sin(x)$, so that the analytical results mentioned are applicable.
Figure~\ref{fig:ex2} shows the case of noisy oscillators. Figure~\ref{fig:ex3} shows the case of deterministic oscillators with the
Cauchy distribution of the natural frequencies (the width parameter of the Cauchy distribution is $\Delta\!=\!1$).
In both cases, the dynamics of a system with random phase shifts nicely corresponds to the analytically predicted behavior of the reduced system.
For the deterministic case, there is an additional (to the comparison of the order parameter with the theoretical prediction) indicator for the validity of the effective coupling.
As it follows from the Ott-Antonsen~\cite{Ott-Antonsen-08} solution of the problem,
the distribution of the phases in the thermodynamic limit $N\to\infty$ is a wrapped Cauchy distribution.
The most direct check of this prediction is the calculation of the higher circular cumulants of the distribution~\cite{Tyulkina_etal-18},
which should vanish for the wrapped Cauchy distribution.
We show the absolute values of the second and the third cumulants in Fig.~\ref{fig:ex3}.
Their values are indeed small compared to the first cumulant (which is the Kuramoto order parameter $Z_{1}$),
and this smallness is even improved as the size of the population grows.

\begin{figure}[!t]
\centering
\includegraphics[width=0.5\columnwidth]{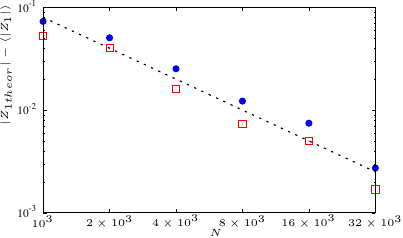}
\caption{Deviations from the theoretical prediction of the main order parameter vs. ensemble size $N$. The same coupling function and phase disorder as in Fig.~\ref{fig:ex3} were used but with an additional disorder in coupling strengths. Red boxes: uniform distribution of $A_{jk}$ in the interval $0\leq A_{jk}\!\leq\!2$; blue circles: $A_{jk}$ takes values $0$ or $2$ with probability $1/2$. The black dotted line shows the law $\!\sim\! N^{-1}$.}
\label{fig:dcs}
\end{figure}

We illustrate the case of disorder in coupling strengths in Fig.~\ref{fig:dcs}. 
As already can be seen in the case of purely phase shifts disorder Fig.~\ref{fig:ex3},
one needs large system sizes $N$ to become closer to the theoretical prediction. This effect is even more pronounced when both the phase shifts and the coupling strengths are random. Therefore, for the same setup as presented in Fig.~\ref{fig:ex3}, we performed calculations for a selected coupling strength $K\!=\!0.625$, for which theory predicts
$|Z_1|\!=\!0.6$. We added disorder in coupling strengths in such a way that $\av{A_{jk}}\!=\!1$, and followed deviations of the obtained values of $\av{|Z_1|}$  from the theoretical prediction. As it follows from Fig.~\ref{fig:dcs}, the deviations decrease with $N$ roughly as $\sim\! N^{-1}$, although they are rather pronounced for small $N\lesssim \!1000$. We tested two distributions of $A_{jk}$, one the uniform in the interval $0\leq A_{jk}\!\leq\! 2$, and one 
bimodal, where $A_{jk}$ takes values $0$ and $2$ with probability $1/2$. In the latter case, finite-size deviations from the theoretical prediction are stronger.

\begin{figure}[!t]
\centering
\includegraphics[width=8.6cm]{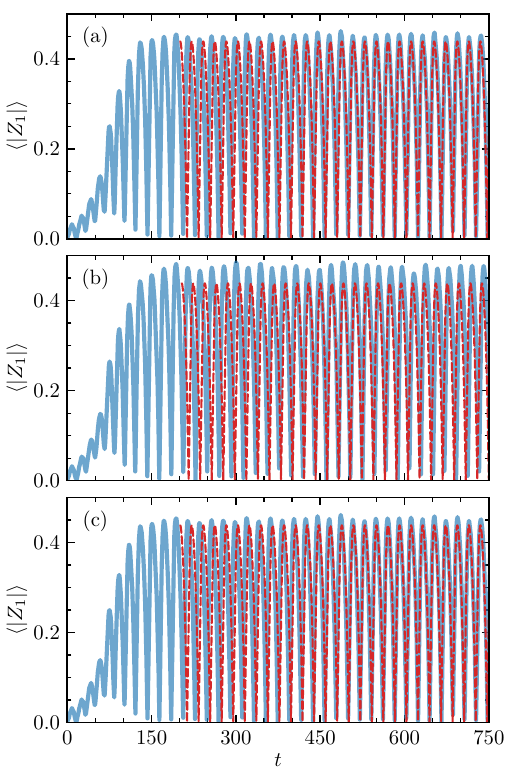}
\caption{Oscillatory partially synchronized regime the noiseless case the Kuramoto-Daido model with two components of the symmetric bimodal distribution having the Cauchy shapes:
$W(\w)=\bigl(W_{C}(\w+\Omega)+W_{C}(\w-\Omega)\bigr)\bigl/2\bigr.$,
where $\Omega=0.25$ and $\Delta=0.125$.
Bold blue solid lines: the dynamics of the order parameter $\av{|Z_{1}|}$ in the systems
of $N=64\times10^{3}$ oscillators with three coupling functions:
panel (a): $F(x)=1.25\sin(x)+0.75\sin(3x)\bigr)$, panel (b): $F(x)=1.25\sin(x)+2.25\sin(2x)\bigr)$, and panel (c): $F(x)=1.25\sin(x)+2.25\sin(3x)\bigr)$.
Thin red dashed curves: the dynamics corresponding to a limit cycle in the ordinary differential equations for 
the two subgroup order parameters, which can be obtained for the Kuramoto model with the effective coupling
$\mathcal{F}(x)=0.625\sin(x)$ using the Ott-Antonsen approach in the thermodynamic limit.}
\label{fig:oscillations}
\end{figure}

Our following example is a system with a bimodal distribution of natural frequencies~\cite{Bonilla_etal-98,Martens-2009,Campa-2020,Kostin-2023}.
Now, the corresponding dynamics may indeed be more complicated than for an unimodal distribution,
with a region of bistability between asynchronous and synchronous states.
A partially synchronous state may be characterized either by a constant or oscillating order parameter.
We  consider the noiseless case of the Kuramoto-Daido model
with several harmonics in the coupling function~\eqref{eq:ff} and with two components of the symmetric bimodal distribution
$W(\w)$ having the Cauchy shapes, i.e. $W(\w)=\bigl(W_{C}(\w+\Omega)+W_{C}(\w-\Omega)\bigr)\bigl/2\bigr.$.
According to our analysis, with the random phase shifts $\alpha_{jk}$ drawn from the one-mode distribution~\eqref{eq:psd} with $M=1$,
the model under consideration can be simplified to a system with an effective coupling function $\mathcal{F}(x)\sim \sin(x)$ that contains only the first harmonic.
The dynamics of the latter oscillator systems with a symmetric bimodal distribution was studied in detail~\cite{Bonilla_etal-98,Martens-2009,Campa-2020,Kostin-2023}.
In particular, in Ref.~\onlinecite{Martens-2009}, possible dynamical regimes and bifurcation between them were comprehensively analyzed
using the macroscopic approaches based on the Ott-Antonsen ansatz.

In Fig.~\ref{fig:oscillations}, we show the oscillatory regimes for the Kuramoto-Daido models with three different coupling functions, each having two harmonics.
However, the effective averaged dynamics is the same, which is clearly seen in the Figure. It is worth mentioning that for good quantitative (not just qualitative) agreement, the number $N$ of ensemble units should be quite large;  we used $N=64\times10^{3}$.

\begin{figure}[!t]
\centering
\includegraphics[width=8.6cm]{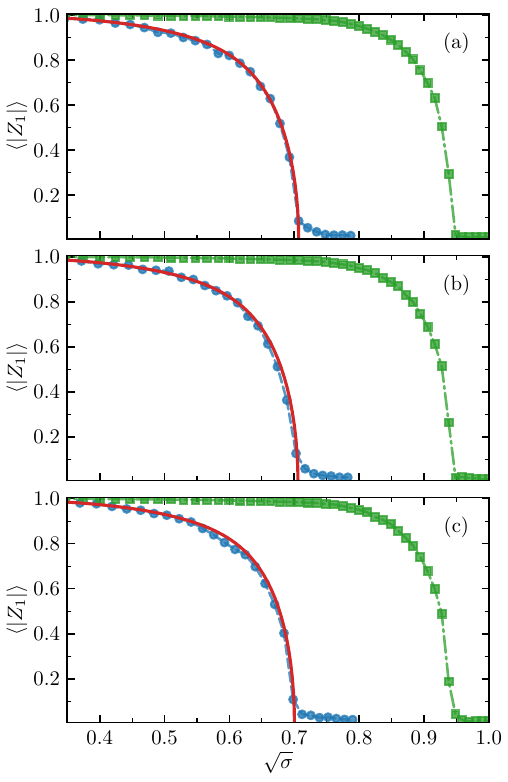}
\caption{ Behavior of the first order parameter $\av{\left|Z_{1}\right|}$ an ensemble of $N\!=\!12\!\times\!10^{3}$ noisy rotators \eqref{eq:rm} with equal natural frequencies ($\w_{k}=\Omega$) and coupling function \eqref{eq:2h} with $K\!=\!1$ in dependence on $\sqrt{\sigma}$ for different values of the moment of inertia $\mu$: $(\mathrm{a})\,\mu=0.02$, $(\mathrm{b})\,\mu=0.1$, and $(\mathrm{c})\,\mu=0.5$. Green squares and blue circles are simulations without and with phase shifts (taken from a distribution \eqref{eq:psd} with $M=1$), respectively. Solid red lines show the theoretical prediction for this order parameter according to Eq.~\eqref{eq:ansolnoise}.}
\label{fig:rotnoise}
\end{figure}

Next, we present simulations of the rotator model~\eqref{eq:rm}.
We consider the case of noisy oscillators with equal natural frequencies and coupling function~\eqref{eq:2h}.
We consider random phase shifts $\alpha_{jk}$ distributed according to~\eqref{eq:psd} with $M=1$. 
Thus, the effective coupling function is $\mathcal{F}(x)=0.5\,K\sin(x)$.
For such coupling, the analytical expression for the order parameter in dependence on the noise intensity $\sigma^{2}$ can be written in the parametric (parameter $R$) form~\cite{Munyaev_etal_2020}
\begin{equation}
|Z_1|=\frac{2\pi R I_0^2(R)I_1(R)}{2\pi R I_0^2(R)+\mu K I_1(R)},\quad
\sigma^2=\frac{K |Z_1|}{2R}\;.
\label{eq:ansolnoise}
\end{equation}
Here, $I_{0}(R)$ and $I_{1}(R)$ are the principal branches of the modified Bessel functions of the first kind with orders $0$ and $1$, respectively.
Relation~\eqref{eq:ansolnoise} is valid for small $\mu$, in the first order in this parameter.
Note, for a vanishing $\mu$, we obtain the equations determining the solid green line in Fig.~\ref{fig:ex2}.
In Fig. \ref{fig:rotnoise}, we compare the dynamics of a noisy ensemble of rotators with and without random phase shifts.
One can see that the effect of the moment of inertia $\mu$ on the coherence level of the established state is relatively small.

\section{Reduction in the case of global random phase shifts}
\label{sec:gps}

\looseness=-1
The theory above was based on the assumption of independence of the phases and the phase shifts, which is plausible if there are many different phase shifts for a given oscillator, like for random pairwise phase shifts. Here, we consider a slightly different setup, where phase shifts $\alpha_{jk}$ are not attributed to connections $j\to k$, but separately to the driven or to the driving system; we call such a situation ``global random phase shifts''. Thus, these phase shifts have one index instead of two. In the case of global phase shifts, the averaging requires additional justification. Such an analysis has been performed in Ref.~\onlinecite{smirnov2024dynamics} for the Kuramoto-Daido setup; here, we present a similar consideration for the Winfree model with noise.

We consider noisy oscillators with global random phase shifts in coupling
\begin{equation}
\dot\vp_k=\sigma\xi_k(t)+\frac{1}{N}S(\vp_k-\alpha_k)\sum_j  Q(\vp_j-\beta_j)\;,
\label{eq:wmp4}
\end{equation}
where $\xi_k(t)$ are independent white Gaussian forces with zero mean $\av{\xi}=0$, and auto-correlation $\av{\xi_k(t')\,\xi_j(t+t')}\!=\!2\delta_{jk}\delta(t)$.
The phase shifts $\alpha_k$ are attributed to the driven systems; they characterize the phase shifts in the phase sensitivity function independently of the driving. On the other hand, the phase shifts $\beta_j$ are attributed to the driving units; they characterize the force functions. Remarkably, by a change of variables $\theta_k=\vp_k-\alpha_k$, one can transfer both shifts to the driving unit: 
\begin{equation}
\dot\theta_k=\sigma\xi_k(t)+\frac{1}{N}S(\theta_k)\sum_j  Q(\theta_j+\alpha_j-\beta_j)
\label{eq:wmp41}
\end{equation}
We thus denote $\gamma_j=\beta_j-\alpha_j$ and consider it as a single combined global 
random phase shit with probability density $\mathit{\Gamma}(\gamma)$.

In the thermodynamic limit $N\to\infty$, we describe the population with the probability density $P(\theta,t|\gamma)$, which generally can depend on the phase shift $\gamma$.
The Fokker-Planck equation for this density reads
\begin{equation}
\frac{\partial P}{\partial t}+\frac{\partial}{\partial\theta}\Bigl(S(\theta)\bar{Q}(t)\Bigr)=\sigma^2\frac{\partial^2P}{\partial \theta^2}\;,
\label{eq:wmp5}
\end{equation}
where
\begin{equation}
\bar{Q}(t)=\!\int_{0}^{2\pi}\!\! d\theta \int_{0}^{2\pi}\!\! d\gamma\, Q(\theta-\gamma) \mathit{\Gamma}(\gamma) P(\theta,t|\gamma)
\label{eq:wmp6}
\end{equation}
contains the averaging over the phases and the phase shifts. The crucial observation is that, although $P(\theta,t|\gamma)$ can potentially depend on $\gamma$, for example, an initial condition at time $t=0$ can contain a $\gamma$-dependence, the equation for $P(\theta,t|\gamma)$ does not contain $\gamma$.
Thus, if this equation, for any prescribed time-dependent function $\bar{Q}(t)$, has a unique attracting solution $P(\theta,t)$ which does not depend on the initial conditions, then the statistical dependence of the phases on the phase shits $\gamma$ disappears, even if it was presented initially.
The property to have a unique solution is sometimes called ``Global Asymptotic Stability''. It is rather natural for the dissipative parabolic equation~\eqref{eq:wmp5} on the circle. Although we found only proofs for time-independent $\bar{Q}$ in the literature, see Refs.~\onlinecite{Gardiner-96,Calogero-12}, it appears that such a proof can be extended to the nonstationary case too~\cite{Z}. For a master equation, the global asymptotic stability has been established in Ref.~\onlinecite{Earnshaw-Keener-10}. 

Taking into account that $P(\theta,t|\gamma)\to P(\theta,t)$ at large times,
we insert the $\gamma$-independent density in Eq.~\eqref{eq:wmp6} and 
obtain
\begin{equation}
\bar{Q}=\!\int_{0}^{2\pi}\!\! d\theta\, P(\theta,t) \mathcal{Q}(\theta),\qquad \mathcal{Q}(\theta)=\!\int_{0}^{2\pi}\!\! d\gamma\, \mathit{\Gamma}(\gamma)Q(\theta-\gamma)\,.
\label{eq:wmp7}
\end{equation}
We see that the forcing function reduces to an effective phase-shift-independent forcing function $\mathcal{Q}(\theta)$, which is the convolution of the original coupling function and the distribution of the phase shifts. This result parallels expression \eqref{eq:wmav}. A similar expression holds for the Kuramoto-Daido coupling (cf.~Ref.~\onlinecite{smirnov2024dynamics}).

\begin{figure}[!t]
\centering
\includegraphics[width=8.6cm]{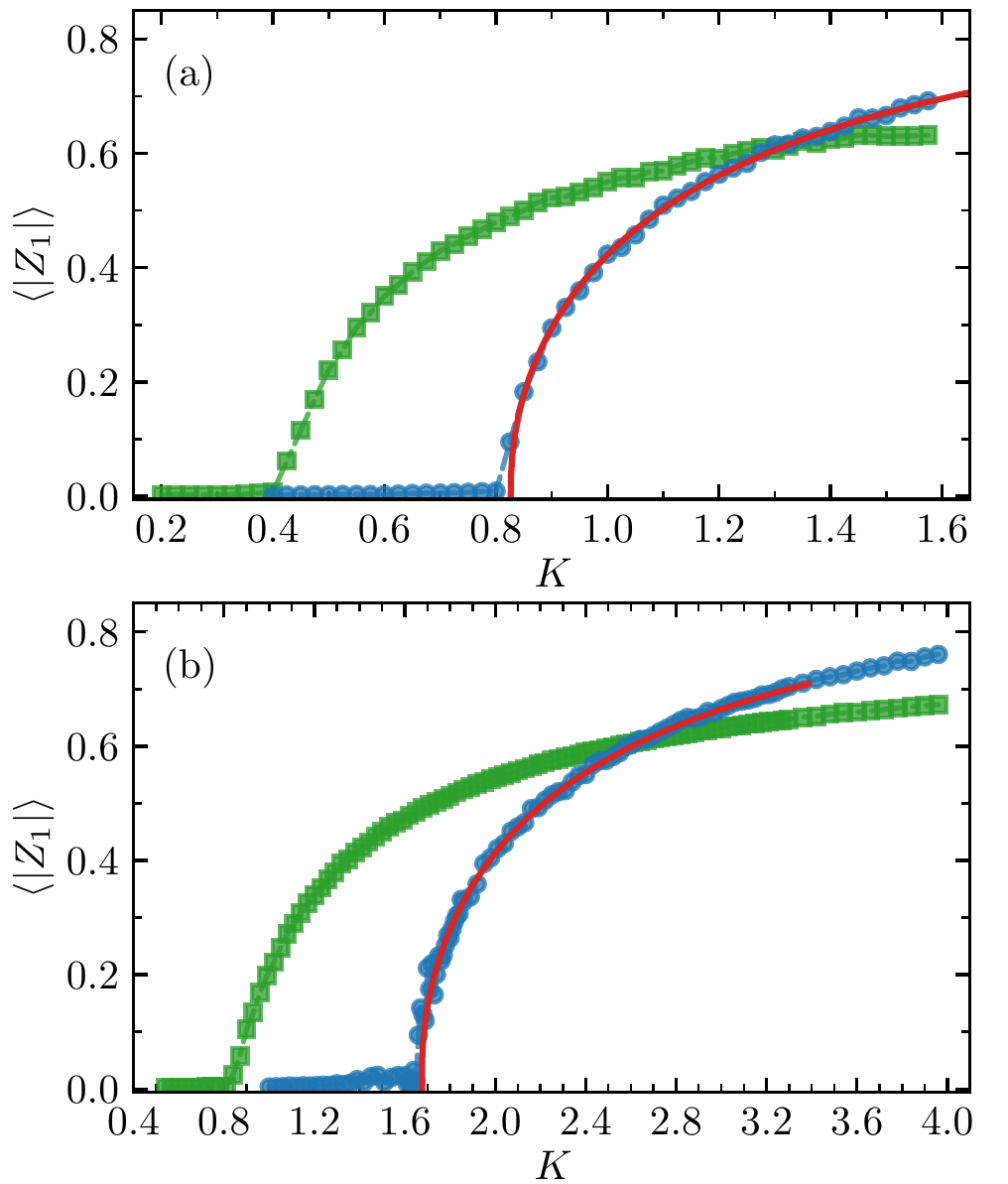} 
\caption{The average values of the order parameter $\av{|Z_1|}$ vs the parameter $K$ of the coupling function $F(x)=K\bigl(\sin(x)-1.5\sin(2x)+0.5\cos(2x)\bigr)$ for two values of the width parameter $\Delta$ of the Cauchy distribution~\eqref{eq:cauchy} of the frequencies $\omega_{k}$: (a) $\Delta=0.2$ and (b) $\Delta=0.4$. Green boxes: without phase shifts; blue circles: with phase shifts. The red lines show the theoretical prediction for the order parameter $\left|Z_{1}\right|$ according to Ref.~\onlinecite{Munyaev_etal_2020}. System size $N=48\times 10^4$.}
\label{fig:rotnoise2}
\end{figure}

In the  example Fig.~\ref{fig:rotnoise2} we consider for the coupled rotators~\eqref{eq:rm} (parameter of inertia $\mu\!=\!0.1$),
we include both a Cauchy distribution of natural frequencies and noise with amplitude $\sigma\!=\!0.05$.
Here, we take the global phase shifts according to $F(\vp_j-\vp_k-\alpha_j)$.
In Fig.~\ref{fig:rotnoise2}, we choose the coupling function in the form $F(x)=K\bigl(\sin(x)-1.5\sin(2x)+0.5\cos(2x)\bigr)$
and present the average values of the order parameter $\av{|Z_1|}$ for simulations without phase shifts
and for phase shifts distributed according to \eqref{eq:psd} with $M=1$.
One can see that the latter data nicely fits the theoretical prediction of Ref.~\onlinecite{Munyaev_etal_2020}.

\section{Conclusion}
\label{sec:conc}

We first summarize the results of this paper. We have considered different models of globally coupled phase oscillators and rotators. In the case of a ``maximal disorder'', all the coupling functions are distinct and random, sampled from some distribution.  Based on the assumption of independence of the phases and the coupling functions in the thermodynamic limit, we derived the averaged equations for the phases, where effective deterministic coupling functions enter. A more detailed consideration was devoted to the case where the shapes of the random coupling functions are the same, but the amplitudes and the phase shifts are random. Then, the effective functions are renormalized convolutions of the original coupling functions and the distribution densities of the phase shifts.  In the Fourier representation, the Fourier modes of the coupling functions are multiplied by Fourier modes of the distribution densities of the phase shifts. This means that the effective averaged coupling is ``simpler'' than the original one. In particular, if the distribution of the phase shifts possesses just one Fourier mode, the effective coupling function will possess only this mode, too. This property allows us to check the validity of the approach numerically because, for the one-mode coupling function, there are theoretical predictions for the behavior of the order parameters.

A special case is a maximally frustrating one, where the averaged coupling function vanishes. For the coupling strengths disorder, this happens in the Daido model~\cite{Daido-92}; in the case of random phase shifts, this occurs for a uniform in the range $0\leq\alpha_{jk}\leq 2\pi$ distribution of the phase shifts. Our theory predicts that in the thermodynamic limit the interaction vanishes. However, certain synchronization phenomena can still be observed in finite ensembles with random phase shifts, as demonstrated recently in Ref.~\onlinecite{pikovsky2024dynamics}. 
We also expect that for other distributions there can be pronounced deviations from the averaged behavior for finite ensembles. This issue definitely deserves further exploration.

\acknowledgments 
L.A.S. acknowledges support from the Russian Science Foundation (grant no. 22-12-00348) and the Ministry of Science and Education of Russian Federation (project no. FSWR-2024-0005).

\section*{Data availability}
All numerical experiments are described in the paper and can be reproduced without additional information.

\bibliography{bibliography}

\begin{thebibliography}{35}%
\makeatletter
\providecommand \@ifxundefined [1]{%
 \@ifx{#1\undefined}
}%
\providecommand \@ifnum [1]{%
 \ifnum #1\expandafter \@firstoftwo
 \else \expandafter \@secondoftwo
 \fi
}%
\providecommand \@ifx [1]{%
 \ifx #1\expandafter \@firstoftwo
 \else \expandafter \@secondoftwo
 \fi
}%
\providecommand \natexlab [1]{#1}%
\providecommand \enquote  [1]{``#1''}%
\providecommand \bibnamefont  [1]{#1}%
\providecommand \bibfnamefont [1]{#1}%
\providecommand \citenamefont [1]{#1}%
\providecommand \href@noop [0]{\@secondoftwo}%
\providecommand \href [0]{\begingroup \@sanitize@url \@href}%
\providecommand \@href[1]{\@@startlink{#1}\@@href}%
\providecommand \@@href[1]{\endgroup#1\@@endlink}%
\providecommand \@sanitize@url [0]{\catcode `\\12\catcode `\$12\catcode
  `\&12\catcode `\#12\catcode `\^12\catcode `\_12\catcode `\%12\relax}%
\providecommand \@@startlink[1]{}%
\providecommand \@@endlink[0]{}%
\providecommand \url  [0]{\begingroup\@sanitize@url \@url }%
\providecommand \@url [1]{\endgroup\@href {#1}{\urlprefix }}%
\providecommand \urlprefix  [0]{URL }%
\providecommand \Eprint [0]{\href }%
\providecommand \doibase [0]{https://doi.org/}%
\providecommand \selectlanguage [0]{\@gobble}%
\providecommand \bibinfo  [0]{\@secondoftwo}%
\providecommand \bibfield  [0]{\@secondoftwo}%
\providecommand \translation [1]{[#1]}%
\providecommand \BibitemOpen [0]{}%
\providecommand \bibitemStop [0]{}%
\providecommand \bibitemNoStop [0]{.\EOS\space}%
\providecommand \EOS [0]{\spacefactor3000\relax}%
\providecommand \BibitemShut  [1]{\csname bibitem#1\endcsname}%
\let\auto@bib@innerbib\@empty
\bibitem [{\citenamefont {Winfree}(1967)}]{Winfree-67}%
  \BibitemOpen
  \bibfield  {author} {\bibinfo {author} {\bibfnamefont {A.~T.}\ \bibnamefont
  {Winfree}},\ }\bibfield  {title} {\enquote {\bibinfo {title} {Biological
  rhythms and the behavior of populations of coupled oscillators},}\
  }\href@noop {} {\bibfield  {journal} {\bibinfo  {journal} {J. Theor. Biol.}\
  }\textbf {\bibinfo {volume} {16}},\ \bibinfo {pages} {15--42} (\bibinfo
  {year} {1967})}\BibitemShut {NoStop}%
\bibitem [{\citenamefont {Kuramoto}(1975)}]{Kuramoto-75}%
  \BibitemOpen
  \bibfield  {author} {\bibinfo {author} {\bibfnamefont {Y.}~\bibnamefont
  {Kuramoto}},\ }\bibfield  {title} {\enquote {\bibinfo {title}
  {Self-entrainment of a population of coupled nonlinear oscillators},}\ }in\
  \href@noop {} {\emph {\bibinfo {booktitle} {International Symposium on
  Mathematical Problems in Theoretical Physics}}},\ \bibinfo {editor} {edited
  by\ \bibinfo {editor} {\bibfnamefont {H.}~\bibnamefont {Araki}}}\ (\bibinfo
  {publisher} {Springer Lecture Notes Phys., v. 39},\ \bibinfo {address} {New
  York},\ \bibinfo {year} {1975})\ p.\ \bibinfo {pages} {420}\BibitemShut
  {NoStop}%
\bibitem [{\citenamefont {Strogatz}\ and\ \citenamefont
  {Stewart}(1993)}]{Strogatz-Stewart-93}%
  \BibitemOpen
  \bibfield  {author} {\bibinfo {author} {\bibfnamefont {S.~H.}\ \bibnamefont
  {Strogatz}}\ and\ \bibinfo {author} {\bibfnamefont {I.}~\bibnamefont
  {Stewart}},\ }\bibfield  {title} {\enquote {\bibinfo {title} {Coupled
  oscillators and biological synchronization},}\ }\href@noop {} {\bibfield
  {journal} {\bibinfo  {journal} {Scientific American}\ ,\ \bibinfo {pages}
  {68--75}} (\bibinfo {year} {1993})}\BibitemShut {NoStop}%
\bibitem [{\citenamefont {Pikovsky}, \citenamefont {Rosenblum},\ and\
  \citenamefont {Kurths}(2001)}]{Pikovsky-Rosenblum-Kurths-01}%
  \BibitemOpen
  \bibfield  {author} {\bibinfo {author} {\bibfnamefont {A.}~\bibnamefont
  {Pikovsky}}, \bibinfo {author} {\bibfnamefont {M.}~\bibnamefont
  {Rosenblum}},\ and\ \bibinfo {author} {\bibfnamefont {J.}~\bibnamefont
  {Kurths}},\ }\href@noop {} {\emph {\bibinfo {title} {Synchronization. A
  Universal Concept in Nonlinear Sciences.}}}\ (\bibinfo  {publisher}
  {Cambridge University Press},\ \bibinfo {address} {Cambridge},\ \bibinfo
  {year} {2001})\BibitemShut {NoStop}%
\bibitem [{\citenamefont {Glass}\ and\ \citenamefont
  {Mackey}(1988)}]{Glass-Mackey-88}%
  \BibitemOpen
  \bibfield  {author} {\bibinfo {author} {\bibfnamefont {L.}~\bibnamefont
  {Glass}}\ and\ \bibinfo {author} {\bibfnamefont {M.~C.}\ \bibnamefont
  {Mackey}},\ }\href@noop {} {\emph {\bibinfo {title} {From Clocks to Chaos:
  {T}he Rhythms of Life.}}}\ (\bibinfo  {publisher} {Princeton Univ. Press},\
  \bibinfo {address} {Princeton, NJ},\ \bibinfo {year} {1988})\BibitemShut
  {NoStop}%
\bibitem [{\citenamefont {Buzs{\'a}ki}(2006)}]{Buzsaki-06}%
  \BibitemOpen
  \bibfield  {author} {\bibinfo {author} {\bibfnamefont {G.}~\bibnamefont
  {Buzs{\'a}ki}},\ }\href@noop {} {\emph {\bibinfo {title} {Rhythms of the
  brain}}}\ (\bibinfo  {publisher} {Oxford UP},\ \bibinfo {address} {Oxford},\
  \bibinfo {year} {2006})\BibitemShut {NoStop}%
\bibitem [{\citenamefont {Daffertshofer}\ and\ \citenamefont
  {Pietras}(2020)}]{daffertshofer2020phase}%
  \BibitemOpen
  \bibfield  {author} {\bibinfo {author} {\bibfnamefont {A.}~\bibnamefont
  {Daffertshofer}}\ and\ \bibinfo {author} {\bibfnamefont {B.}~\bibnamefont
  {Pietras}},\ }\bibfield  {title} {\enquote {\bibinfo {title} {Phase
  synchronization in neural systems},}\ }\href@noop {} {\bibfield  {journal}
  {\bibinfo  {journal} {Synergetics}\ ,\ \bibinfo {pages} {221--233}} (\bibinfo
  {year} {2020})}\BibitemShut {NoStop}%
\bibitem [{\citenamefont {Kalloniatis}(2010)}]{kalloniatis2010incoherence}%
  \BibitemOpen
  \bibfield  {author} {\bibinfo {author} {\bibfnamefont {A.~C.}\ \bibnamefont
  {Kalloniatis}},\ }\bibfield  {title} {\enquote {\bibinfo {title} {From
  incoherence to synchronicity in the network {K}uramoto model},}\ }\href@noop
  {} {\bibfield  {journal} {\bibinfo  {journal} {Physical Review E}\ }\textbf
  {\bibinfo {volume} {82}},\ \bibinfo {pages} {066202} (\bibinfo {year}
  {2010})}\BibitemShut {NoStop}%
\bibitem [{\citenamefont {Chiba}, \citenamefont {Medvedev},\ and\ \citenamefont
  {Mizuhara}(2018)}]{chiba2018bifurcations}%
  \BibitemOpen
  \bibfield  {author} {\bibinfo {author} {\bibfnamefont {H.}~\bibnamefont
  {Chiba}}, \bibinfo {author} {\bibfnamefont {G.~S.}\ \bibnamefont
  {Medvedev}},\ and\ \bibinfo {author} {\bibfnamefont {M.~S.}\ \bibnamefont
  {Mizuhara}},\ }\bibfield  {title} {\enquote {\bibinfo {title} {Bifurcations
  in the {K}uramoto model on graphs},}\ }\href@noop {} {\bibfield  {journal}
  {\bibinfo  {journal} {Chaos: An Interdisciplinary Journal of Nonlinear
  Science}\ }\textbf {\bibinfo {volume} {28}} (\bibinfo {year}
  {2018})}\BibitemShut {NoStop}%
\bibitem [{\citenamefont {Juh{\'a}sz}, \citenamefont {Kelling},\ and\
  \citenamefont {{\'O}dor}(2019)}]{juhasz2019critical}%
  \BibitemOpen
  \bibfield  {author} {\bibinfo {author} {\bibfnamefont {R.}~\bibnamefont
  {Juh{\'a}sz}}, \bibinfo {author} {\bibfnamefont {J.}~\bibnamefont
  {Kelling}},\ and\ \bibinfo {author} {\bibfnamefont {G.}~\bibnamefont
  {{\'O}dor}},\ }\bibfield  {title} {\enquote {\bibinfo {title} {Critical
  dynamics of the {K}uramoto model on sparse random networks},}\ }\href@noop {}
  {\bibfield  {journal} {\bibinfo  {journal} {Journal of Statistical Mechanics:
  Theory and Experiment}\ }\textbf {\bibinfo {volume} {2019}},\ \bibinfo
  {pages} {053403} (\bibinfo {year} {2019})}\BibitemShut {NoStop}%
\bibitem [{\citenamefont {Park}, \citenamefont {Rhee},\ and\ \citenamefont
  {Choi}(1998)}]{Park-Rhee-Choi-98}%
  \BibitemOpen
  \bibfield  {author} {\bibinfo {author} {\bibfnamefont {K.}~\bibnamefont
  {Park}}, \bibinfo {author} {\bibfnamefont {S.~W.}\ \bibnamefont {Rhee}},\
  and\ \bibinfo {author} {\bibfnamefont {M.~Y.}\ \bibnamefont {Choi}},\
  }\bibfield  {title} {\enquote {\bibinfo {title} {Glass synchronization in the
  network of oscillators with random phase shifts},}\ }\href
  {https://doi.org/10.1103/PhysRevE.57.5030} {\bibfield  {journal} {\bibinfo
  {journal} {Phys. Rev. E}\ }\textbf {\bibinfo {volume} {57}},\ \bibinfo
  {pages} {5030--5035} (\bibinfo {year} {1998})}\BibitemShut {NoStop}%
\bibitem [{\citenamefont {Smirnov}\ and\ \citenamefont
  {Pikovsky}(2024)}]{smirnov2024dynamics}%
  \BibitemOpen
  \bibfield  {author} {\bibinfo {author} {\bibfnamefont {L.~A.}\ \bibnamefont
  {Smirnov}}\ and\ \bibinfo {author} {\bibfnamefont {A.}~\bibnamefont
  {Pikovsky}},\ }\bibfield  {title} {\enquote {\bibinfo {title} {Dynamics of
  oscillator populations globally coupled with distributed phase shifts},}\
  }\href@noop {} {\bibfield  {journal} {\bibinfo  {journal} {Phys. Rev. Lett.}\
  }\textbf {\bibinfo {volume} {132}},\ \bibinfo {pages} {107401} (\bibinfo
  {year} {2024})}\BibitemShut {NoStop}%
\bibitem [{\citenamefont {Shinomoto}\ and\ \citenamefont
  {Kuramoto}(1986)}]{Shinomoto-Kuramoto-86}%
  \BibitemOpen
  \bibfield  {author} {\bibinfo {author} {\bibfnamefont {S.}~\bibnamefont
  {Shinomoto}}\ and\ \bibinfo {author} {\bibfnamefont {Y.}~\bibnamefont
  {Kuramoto}},\ }\bibfield  {title} {\enquote {\bibinfo {title} {Phase
  transitions in active rotator systems},}\ }\href@noop {} {\bibfield
  {journal} {\bibinfo  {journal} {Prog. Theor. Phys.}\ }\textbf {\bibinfo
  {volume} {75}},\ \bibinfo {pages} {1105--1110} (\bibinfo {year}
  {1986})}\BibitemShut {NoStop}%
\bibitem [{\citenamefont {Sakaguchi}, \citenamefont {Shinomoto},\ and\
  \citenamefont {Kuramoto}(1988)}]{Sakaguchi-Shinomoto-Kuramoto-88}%
  \BibitemOpen
  \bibfield  {author} {\bibinfo {author} {\bibfnamefont {H.}~\bibnamefont
  {Sakaguchi}}, \bibinfo {author} {\bibfnamefont {S.}~\bibnamefont
  {Shinomoto}},\ and\ \bibinfo {author} {\bibfnamefont {Y.}~\bibnamefont
  {Kuramoto}},\ }\bibfield  {title} {\enquote {\bibinfo {title} {Phase
  transitions and their bifurcation analysis in a large population of active
  rotators with mean-field coupling},}\ }\href@noop {} {\bibfield  {journal}
  {\bibinfo  {journal} {Prog. Theor. Phys.}\ }\textbf {\bibinfo {volume}
  {79}},\ \bibinfo {pages} {600--607} (\bibinfo {year} {1988})}\BibitemShut
  {NoStop}%
\bibitem [{\citenamefont {Klinshov}\ \emph {et~al.}(2021)\citenamefont
  {Klinshov}, \citenamefont {Kirillov}, \citenamefont {Nekorkin},\ and\
  \citenamefont {Wolfrum}}]{Klinshov2021noise}%
  \BibitemOpen
  \bibfield  {author} {\bibinfo {author} {\bibfnamefont {V.~V.}\ \bibnamefont
  {Klinshov}}, \bibinfo {author} {\bibfnamefont {S.~Y.}\ \bibnamefont
  {Kirillov}}, \bibinfo {author} {\bibfnamefont {V.~I.}\ \bibnamefont
  {Nekorkin}},\ and\ \bibinfo {author} {\bibfnamefont {M.}~\bibnamefont
  {Wolfrum}},\ }\bibfield  {title} {\enquote {\bibinfo {title} {Noise-induced
  dynamical regimes in a system of globally coupled excitable units},}\
  }\href@noop {} {\bibfield  {journal} {\bibinfo  {journal} {Chaos: An
  Interdisciplinary Journal of Nonlinear Science}\ }\textbf {\bibinfo {volume}
  {31}} (\bibinfo {year} {2021})}\BibitemShut {NoStop}%
\bibitem [{\citenamefont {Tanaka}, \citenamefont {Lichtenberg},\ and\
  \citenamefont {Oishi}(1997)}]{Tanaka-Lichtenberg-Oishi-97}%
  \BibitemOpen
  \bibfield  {author} {\bibinfo {author} {\bibfnamefont {H.}~\bibnamefont
  {Tanaka}}, \bibinfo {author} {\bibfnamefont {A.}~\bibnamefont
  {Lichtenberg}},\ and\ \bibinfo {author} {\bibfnamefont {S.}~\bibnamefont
  {Oishi}},\ }\bibfield  {title} {\enquote {\bibinfo {title} {First order phase
  transition resulting from finite inertia in coupled oscillator systems},}\
  }\href@noop {} {\bibfield  {journal} {\bibinfo  {journal} {Phys. Rev. Lett.}\
  }\textbf {\bibinfo {volume} {78}},\ \bibinfo {pages} {2104--2107} (\bibinfo
  {year} {1997})}\BibitemShut {NoStop}%
\bibitem [{\citenamefont {Hong}\ \emph {et~al.}(1999)\citenamefont {Hong},
  \citenamefont {Choi}, \citenamefont {Yi},\ and\ \citenamefont
  {Soh}}]{Hong-Choi-Yi-Soh-99}%
  \BibitemOpen
  \bibfield  {author} {\bibinfo {author} {\bibfnamefont {H.}~\bibnamefont
  {Hong}}, \bibinfo {author} {\bibfnamefont {M.~Y.}\ \bibnamefont {Choi}},
  \bibinfo {author} {\bibfnamefont {J.}~\bibnamefont {Yi}},\ and\ \bibinfo
  {author} {\bibfnamefont {K.-S.}\ \bibnamefont {Soh}},\ }\bibfield  {title}
  {\enquote {\bibinfo {title} {Inertia effects on periodic synchronization in a
  system of coupled oscillators},}\ }\href@noop {} {\bibfield  {journal}
  {\bibinfo  {journal} {Phys. Rev. E}\ }\textbf {\bibinfo {volume} {59}},\
  \bibinfo {pages} {353--363} (\bibinfo {year} {1999})}\BibitemShut {NoStop}%
\bibitem [{\citenamefont {Munyaev}\ \emph {et~al.}(2020)\citenamefont
  {Munyaev}, \citenamefont {Smirnov}, \citenamefont {Kostin}, \citenamefont
  {Osipov},\ and\ \citenamefont {Pikovsky}}]{Munyaev_etal_2020}%
  \BibitemOpen
  \bibfield  {author} {\bibinfo {author} {\bibfnamefont {V.~O.}\ \bibnamefont
  {Munyaev}}, \bibinfo {author} {\bibfnamefont {L.~A.}\ \bibnamefont
  {Smirnov}}, \bibinfo {author} {\bibfnamefont {V.~A.}\ \bibnamefont {Kostin}},
  \bibinfo {author} {\bibfnamefont {G.~V.}\ \bibnamefont {Osipov}},\ and\
  \bibinfo {author} {\bibfnamefont {A.}~\bibnamefont {Pikovsky}},\ }\bibfield
  {title} {\enquote {\bibinfo {title} {Analytical approach to synchronous
  states of globally coupled noisy rotators},}\ }\href@noop {} {\bibfield
  {journal} {\bibinfo  {journal} {New Journal of Physics}\ }\textbf {\bibinfo
  {volume} {22}},\ \bibinfo {pages} {023036} (\bibinfo {year}
  {2020})}\BibitemShut {NoStop}%
\bibitem [{\citenamefont {Munyayev}\ \emph {et~al.}(2023)\citenamefont
  {Munyayev}, \citenamefont {Bolotov}, \citenamefont {Smirnov}, \citenamefont
  {Osipov},\ and\ \citenamefont {Belykh}}]{Munyayev_etal_2023}%
  \BibitemOpen
  \bibfield  {author} {\bibinfo {author} {\bibfnamefont {V.~O.}\ \bibnamefont
  {Munyayev}}, \bibinfo {author} {\bibfnamefont {M.~I.}\ \bibnamefont
  {Bolotov}}, \bibinfo {author} {\bibfnamefont {L.~A.}\ \bibnamefont
  {Smirnov}}, \bibinfo {author} {\bibfnamefont {G.~V.}\ \bibnamefont
  {Osipov}},\ and\ \bibinfo {author} {\bibfnamefont {I.}~\bibnamefont
  {Belykh}},\ }\bibfield  {title} {\enquote {\bibinfo {title} {Cyclops states
  in repulsive kuramoto networks: The role of higher-order coupling},}\ }\href
  {https://doi.org/10.1103/PhysRevLett.130.107201} {\bibfield  {journal}
  {\bibinfo  {journal} {Phys. Rev. Lett.}\ }\textbf {\bibinfo {volume} {130}},\
  \bibinfo {pages} {107201} (\bibinfo {year} {2023})}\BibitemShut {NoStop}%
\bibitem [{\citenamefont {Filatrella}, \citenamefont {Nielsen},\ and\
  \citenamefont {Pedersen}(2008)}]{Filatrella_etall-08}%
  \BibitemOpen
  \bibfield  {author} {\bibinfo {author} {\bibfnamefont {G.}~\bibnamefont
  {Filatrella}}, \bibinfo {author} {\bibfnamefont {A.~H.}\ \bibnamefont
  {Nielsen}},\ and\ \bibinfo {author} {\bibfnamefont {N.~F.}\ \bibnamefont
  {Pedersen}},\ }\bibfield  {title} {\enquote {\bibinfo {title} {Analysis of a
  power grid using a {K}uramoto-like model},}\ }\href@noop {} {\bibfield
  {journal} {\bibinfo  {journal} {Eur. Phys. J. B}\ }\textbf {\bibinfo {volume}
  {61}},\ \bibinfo {pages} {485--491} (\bibinfo {year} {2008})}\BibitemShut
  {NoStop}%
\bibitem [{\citenamefont {Dorfler}\ and\ \citenamefont
  {Bullo}(2012)}]{dorfler2012synchronization}%
  \BibitemOpen
  \bibfield  {author} {\bibinfo {author} {\bibfnamefont {F.}~\bibnamefont
  {Dorfler}}\ and\ \bibinfo {author} {\bibfnamefont {F.}~\bibnamefont
  {Bullo}},\ }\bibfield  {title} {\enquote {\bibinfo {title} {Synchronization
  and transient stability in power networks and nonuniform {K}uramoto
  oscillators},}\ }\href@noop {} {\bibfield  {journal} {\bibinfo  {journal}
  {SIAM Journal on Control and Optimization}\ }\textbf {\bibinfo {volume}
  {50}},\ \bibinfo {pages} {1616--1642} (\bibinfo {year} {2012})}\BibitemShut
  {NoStop}%
\bibitem [{\citenamefont {Kuramoto}(1984)}]{Kuramoto-84}%
  \BibitemOpen
  \bibfield  {author} {\bibinfo {author} {\bibfnamefont {Y.}~\bibnamefont
  {Kuramoto}},\ }\href@noop {} {\emph {\bibinfo {title} {Chemical Oscillations,
  Waves and Turbulence}}}\ (\bibinfo  {publisher} {Springer},\ \bibinfo
  {address} {Berlin},\ \bibinfo {year} {1984})\BibitemShut {NoStop}%
\bibitem [{\citenamefont {Izhikevich}(1998)}]{Izhikevich-98}%
  \BibitemOpen
  \bibfield  {author} {\bibinfo {author} {\bibfnamefont {E.~M.}\ \bibnamefont
  {Izhikevich}},\ }\bibfield  {title} {\enquote {\bibinfo {title} {Phase models
  with explicit time delays},}\ }\href@noop {} {\bibfield  {journal} {\bibinfo
  {journal} {Phys. Rev. E}\ }\textbf {\bibinfo {volume} {58}},\ \bibinfo
  {pages} {905--908} (\bibinfo {year} {1998})}\BibitemShut {NoStop}%
\bibitem [{\citenamefont {Ott}\ and\ \citenamefont
  {Antonsen}(2008)}]{Ott-Antonsen-08}%
  \BibitemOpen
  \bibfield  {author} {\bibinfo {author} {\bibfnamefont {E.}~\bibnamefont
  {Ott}}\ and\ \bibinfo {author} {\bibfnamefont {T.~M.}\ \bibnamefont
  {Antonsen}},\ }\bibfield  {title} {\enquote {\bibinfo {title} {Low
  dimensional behavior of large systems of globally coupled oscillators},}\
  }\href@noop {} {\bibfield  {journal} {\bibinfo  {journal} {CHAOS}\ }\textbf
  {\bibinfo {volume} {18}},\ \bibinfo {pages} {037113} (\bibinfo {year}
  {2008})}\BibitemShut {NoStop}%
\bibitem [{\citenamefont {Tyulkina}\ \emph {et~al.}(2018)\citenamefont
  {Tyulkina}, \citenamefont {Goldobin}, \citenamefont {Klimenko},\ and\
  \citenamefont {Pikovsky}}]{Tyulkina_etal-18}%
  \BibitemOpen
  \bibfield  {author} {\bibinfo {author} {\bibfnamefont {I.~V.}\ \bibnamefont
  {Tyulkina}}, \bibinfo {author} {\bibfnamefont {D.~S.}\ \bibnamefont
  {Goldobin}}, \bibinfo {author} {\bibfnamefont {L.~S.}\ \bibnamefont
  {Klimenko}},\ and\ \bibinfo {author} {\bibfnamefont {A.}~\bibnamefont
  {Pikovsky}},\ }\bibfield  {title} {\enquote {\bibinfo {title} {Dynamics of
  noisy oscillator populations beyond the {O}tt-{A}ntonsen ansatz},}\
  }\href@noop {} {\bibfield  {journal} {\bibinfo  {journal} {Phys. Rev. Lett.}\
  }\textbf {\bibinfo {volume} {120}},\ \bibinfo {pages} {264101} (\bibinfo
  {year} {2018})}\BibitemShut {NoStop}%
\bibitem [{\citenamefont {Bonilla}, \citenamefont {Vicente},\ and\
  \citenamefont {Spigler}(1998)}]{Bonilla_etal-98}%
  \BibitemOpen
  \bibfield  {author} {\bibinfo {author} {\bibfnamefont {L.~L.}\ \bibnamefont
  {Bonilla}}, \bibinfo {author} {\bibfnamefont {C.~J.~P.}\ \bibnamefont
  {Vicente}},\ and\ \bibinfo {author} {\bibfnamefont {R.}~\bibnamefont
  {Spigler}},\ }\bibfield  {title} {\enquote {\bibinfo {title} {Time-periodic
  phases in populations of nonlinearly coupled oscillators with bimodal
  frequency distributions},}\ }\href@noop {} {\bibfield  {journal} {\bibinfo
  {journal} {Physica D: Nonlinear Phenomena}\ }\textbf {\bibinfo {volume}
  {113}},\ \bibinfo {pages} {79 -- 97} (\bibinfo {year} {1998})}\BibitemShut
  {NoStop}%
\bibitem [{\citenamefont {Martens}\ \emph {et~al.}(2009)\citenamefont
  {Martens}, \citenamefont {Barreto}, \citenamefont {Strogatz}, \citenamefont
  {Ott}, \citenamefont {So},\ and\ \citenamefont {Antonsen}}]{Martens-2009}%
  \BibitemOpen
  \bibfield  {author} {\bibinfo {author} {\bibfnamefont {E.~A.}\ \bibnamefont
  {Martens}}, \bibinfo {author} {\bibfnamefont {E.}~\bibnamefont {Barreto}},
  \bibinfo {author} {\bibfnamefont {S.~H.}\ \bibnamefont {Strogatz}}, \bibinfo
  {author} {\bibfnamefont {E.}~\bibnamefont {Ott}}, \bibinfo {author}
  {\bibfnamefont {P.}~\bibnamefont {So}},\ and\ \bibinfo {author}
  {\bibfnamefont {T.~M.}\ \bibnamefont {Antonsen}},\ }\bibfield  {title}
  {\enquote {\bibinfo {title} {Exact results for the {K}uramoto model with a
  bimodal frequency distribution},}\ }\href
  {https://doi.org/10.1103/PhysRevE.79.026204} {\bibfield  {journal} {\bibinfo
  {journal} {Phys. Rev. E}\ }\textbf {\bibinfo {volume} {79}},\ \bibinfo
  {pages} {026204} (\bibinfo {year} {2009})}\BibitemShut {NoStop}%
\bibitem [{\citenamefont {Campa}(2020)}]{Campa-2020}%
  \BibitemOpen
  \bibfield  {author} {\bibinfo {author} {\bibfnamefont {A.}~\bibnamefont
  {Campa}},\ }\bibfield  {title} {\enquote {\bibinfo {title} {Phase diagram of
  noisy systems of coupled oscillators with a bimodal frequency
  distribution},}\ }\href {https://doi.org/10.1088/1751-8121/ab79f2} {\bibfield
   {journal} {\bibinfo  {journal} {Journal of Physics A: Mathematical and
  Theoretical}\ }\textbf {\bibinfo {volume} {53}},\ \bibinfo {pages} {154001}
  (\bibinfo {year} {2020})}\BibitemShut {NoStop}%
\bibitem [{\citenamefont {Kostin}\ \emph {et~al.}(2023)\citenamefont {Kostin},
  \citenamefont {Munyaev}, \citenamefont {Osipov},\ and\ \citenamefont
  {Smirnov}}]{Kostin-2023}%
  \BibitemOpen
  \bibfield  {author} {\bibinfo {author} {\bibfnamefont {V.~A.}\ \bibnamefont
  {Kostin}}, \bibinfo {author} {\bibfnamefont {V.~O.}\ \bibnamefont {Munyaev}},
  \bibinfo {author} {\bibfnamefont {G.~V.}\ \bibnamefont {Osipov}},\ and\
  \bibinfo {author} {\bibfnamefont {L.~A.}\ \bibnamefont {Smirnov}},\
  }\bibfield  {title} {\enquote {\bibinfo {title} {{Synchronization transitions
  and sensitivity to asymmetry in the bimodal {K}uramoto systems with {C}auchy
  noise}},}\ }\href {https://doi.org/10.1063/5.0160006} {\bibfield  {journal}
  {\bibinfo  {journal} {Chaos: An Interdisciplinary Journal of Nonlinear
  Science}\ }\textbf {\bibinfo {volume} {33}},\ \bibinfo {pages} {083155}
  (\bibinfo {year} {2023})}\BibitemShut {NoStop}%
\bibitem [{\citenamefont {Gardiner}(1996)}]{Gardiner-96}%
  \BibitemOpen
  \bibfield  {author} {\bibinfo {author} {\bibfnamefont {C.~W.}\ \bibnamefont
  {Gardiner}},\ }\href@noop {} {\emph {\bibinfo {title} {Handbook of Stochastic
  Methods}}}\ (\bibinfo  {publisher} {Springer},\ \bibinfo {address} {Berlin},\
  \bibinfo {year} {1996})\BibitemShut {NoStop}%
\bibitem [{\citenamefont {Calogero}(2012)}]{Calogero-12}%
  \BibitemOpen
  \bibfield  {author} {\bibinfo {author} {\bibfnamefont {S.}~\bibnamefont
  {Calogero}},\ }\bibfield  {title} {\enquote {\bibinfo {title} {Exponential
  convergence to equilibrium for kinetic {F}okker-{P}lanck equations},}\
  }\href@noop {} {\bibfield  {journal} {\bibinfo  {journal} {Comm. Part. Diff.
  Eqs.}\ }\textbf {\bibinfo {volume} {37}},\ \bibinfo {pages} {1357--1390}
  (\bibinfo {year} {2012})}\BibitemShut {NoStop}%
\bibitem [{Z()}]{Z}%
  \BibitemOpen
  \href@noop {} {}\bibinfo {note} {{S}. Zelik (private communication,
  2023)}\BibitemShut {NoStop}%
\bibitem [{\citenamefont {Earnshaw}\ and\ \citenamefont
  {Keener}(2010)}]{Earnshaw-Keener-10}%
  \BibitemOpen
  \bibfield  {author} {\bibinfo {author} {\bibfnamefont {B.~A.}\ \bibnamefont
  {Earnshaw}}\ and\ \bibinfo {author} {\bibfnamefont {J.~P.}\ \bibnamefont
  {Keener}},\ }\bibfield  {title} {\enquote {\bibinfo {title} {Global
  asymptotic stability of solutions of nonautonomous master equations},}\
  }\href@noop {} {\bibfield  {journal} {\bibinfo  {journal} {SIAM Journal on
  Applied Dynamical Systems}\ }\textbf {\bibinfo {volume} {9}},\ \bibinfo
  {pages} {220--237} (\bibinfo {year} {2010})}\BibitemShut {NoStop}%
\bibitem [{\citenamefont {Daido}(1992)}]{Daido-92}%
  \BibitemOpen
  \bibfield  {author} {\bibinfo {author} {\bibfnamefont {H.}~\bibnamefont
  {Daido}},\ }\bibfield  {title} {\enquote {\bibinfo {title} {Quasientrainment
  and slow relaxation in a population of oscillators with random and frustrated
  interactions},}\ }\href@noop {} {\bibfield  {journal} {\bibinfo  {journal}
  {Phys. Rev. Lett.}\ }\textbf {\bibinfo {volume} {68}},\ \bibinfo {pages}
  {1073--1076} (\bibinfo {year} {1992})}\BibitemShut {NoStop}%
\bibitem [{\citenamefont {Pikovsky}\ and\ \citenamefont
  {Bagnoli}(2024)}]{pikovsky2024dynamics}%
  \BibitemOpen
  \bibfield  {author} {\bibinfo {author} {\bibfnamefont {A.}~\bibnamefont
  {Pikovsky}}\ and\ \bibinfo {author} {\bibfnamefont {F.}~\bibnamefont
  {Bagnoli}},\ }\bibfield  {title} {\enquote {\bibinfo {title} {Dynamics of
  oscillator populations with disorder in the coupling phase shifts},}\
  }\href@noop {} {\bibfield  {journal} {\bibinfo  {journal} {New Journal of
  Physics}\ }\textbf {\bibinfo {volume} {26}},\ \bibinfo {pages} {023054}
  (\bibinfo {year} {2024})}\BibitemShut {NoStop}%
\end{thebibliography}%
\end{document}